\documentclass[11pt]{article}
\usepackage{aaspp4}
\begin{document}


\def\kms{km\,s$^{-1}$}
\def\whz{W\,Hz$^{-1}$}
\def\ergcms{erg\,cm$^{-2}$\,s$^{-1}$}
\def\ergsec{erg\,s$^{-1}$}
\def\wisk#1{\ifmmode{#1}\else{$#1$}\fi}
\def\arcpt{\wisk{''\mkern-7.0mu .\mkern1.4mu}}
\def\arcmpt{\wisk{'\mkern-4.0mu .\mkern1.0mu}}
\def\arcdeg{\wisk{^{\circ}\mkern-7.0mu .\mkern1.4mu}}

\newcommand{\stt}{\small\tt}
\newcommand{\wl}{$W_{\lambda}$}
\newcommand{\oiii}{[O {\sc iii}]}
\newcommand{\oii}{[O {\sc ii}]}
\newcommand{\oi}{[O {\sc i}]}
\newcommand{\lya}{Ly\,$\alpha$}
\newcommand{\ha}{H\,$\alpha$}
\newcommand{\hb}{H\,$\beta$}
\newcommand{\civ}{C {\sc iv}}
\newcommand{\nv}{N {\sc v}}
\newcommand{\siv}{Si {\sc iv}}
\newcommand{\mgii}{Mg {\sc ii}}
\newcommand{\feii}{Fe {\sc ii}}
\newcommand{\arad}{$\alpha_{\rm radio}$}
\newcommand{\aopt}{$\alpha_{\rm opt}$}
\newcommand{\gl}{$\lambda$}
\newcommand{\ew}{$W_{\lambda}$}
\newcommand{\reff}{\noindent\hangindent=3em\hangafter=1}

\title{Detection of a CMB decrement towards the $z = 3.8$ quasar pair
PC1643+4631~A~\&~B}

\author{Michael E. Jones\altaffilmark{1}, Richard Saunders\altaffilmark{1},
Joanne C. Baker\altaffilmark{1}, Garret Cotter\altaffilmark{1}, Alastair
Edge\altaffilmark{2}, Keith Grainge\altaffilmark{1}, Toby
Haynes\altaffilmark{1}, Anthony Lasenby\altaffilmark{1}, Guy
Pooley\altaffilmark{1}, Huub R\"ottgering\altaffilmark{3}}

\altaffiltext{1}{Mullard Radio Astronomy
Observatory, Cavendish Laboratory, Madingley Road, Cambridge, CB3 0HE, UK}

\altaffiltext{2}{Institute of Astronomy, Madingley Road,
Cambridge, CB3 0HA, UK}

\altaffiltext{3}{Leiden Observatory, PO Box 9513, 2300 RA, Leiden, The Netherlands}

\begin{abstract} 

In a 15-GHz Ryle-Telescope observation of PC1643+4631 A \& B, a pair of
quasars at redshifts $z = 3.79$ and 3.83 separated by $198\arcsec$ on the sky,
we find a decrement in the cosmic microwave background (CMB) of $-380 \pm
64~\mu$Jy in a $110 \arcsec \times 175 \arcsec$ beam. Assuming this to be a
Sunyaev-Zel'dovich effect due to an intervening cluster, the minimum magnitude
of the central temperature decrement is $560\; \mu$K. A serendipitous ROSAT
observation shows that there is no X-ray-luminous cluster in the direction of
the decrement at $z < 1$. The implied gas mass is $\gtrsim 2 \times 10^{14} \;
{\rm M_{\odot}}$ (assuming a temperature of $\sim 5$ keV), indicating a total
mass of $> 10^{15} \;{\rm M_{\odot}}$. This result demonstrates the existence
of a massive system too distant to be detected by its emission, but which can
be found via its imprint on the CMB.

\end{abstract}


\section{Introduction}

Clusters of galaxies are difficult to identify optically at high redshift
because of the rapid decrease of surface brightness with redshift and
because of confusion with the foreground field. X-ray surveys have successfully detected
clusters out to redshifts $z$ of 0.5--1 (e.g. Gioia et al.  1990), but
are biassed towards finding only the very richest clusters due to the
strong dependence of X-ray luminosity on gas density. The Sunyaev-Zel'dovich
(S-Z) effect (Sunyaev \& Zel'dovich 1972) is a potentially very useful tool
for finding and studying high-redshift clusters because its magnitude is
independent of redshift, with only the angular size of the effect changing
with $z$. Observing techniques have advanced sufficiently that there have now
been detections of the S-Z effect in several clusters (e.g. Birkinshaw et al.
1984; Jones et al. 1993; Wilbanks et al. 1994; Herbig et al. 1995). Our own
S-Z programme with the Ryle Telescope (RT; Jones 1991\nocite{j91}) operating
at 15~GHz has produced S-Z images of several moderately distant ($0.15 < z <
0.55$) clusters (e.g. Grainge et al. 1993; Saunders 1995; Grainge et al.
1996). We report here the results of our initial programme to obtain S-Z
detections of more-distant systems.

\section{Observing Strategy}

We chose to observe the fields of radio-quiet quasars, for two reasons. 
First,
quasars have long been expected to exist in high density regions and
observations of quasar companions support this (e.g. Hu, McMahon \& Egami
1996). Second, high-redshift quasars may serve as markers for intervening
gravitational lenses which magnify them. We drew up a list of northern
quasars at $z > 1$ with a 5-GHz flux density of $< 1$~mJy (in order to
minimize the contamination of any S-Z signal by radio emission from the quasar
itself) and checked that the chosen fields did not contain any known
foreground clusters.  These criteria produced a list of about thirty candidate
quasars, of which three were chosen for initial observation on the basis of
how well their positions would fit in with the rest of the RT observing
programme. The targets were PG0117+213, MS00365 and PC1643+4631A~\&~B; the
radio observations are now described.

\section{Radio observations}

The RT is an nearly east-west, aperture-synthesis telescope operating
at 15~GHz, with eight 13-m-diameter antennas. For this experiment only
five antennas were used, in an array giving projected baselines from
13 to 108 m (0.65--5.4~k$\lambda$) (Grainge et al. 1996). In this
configuration the system temperature of 75~K and bandwidth of 350~MHz
give a sensitivity of $200 \; \mu$Jy in 12~h in a synthesised beam of
FWHM $30\arcsec \times 30\arcsec {\rm cosec} \delta$, with an envelope
beam of $6 \arcmin$ FWHM. For each field, observations of a phase
calibrator are interleaved with the main observation at intervals of
about 20 min, and a primary flux calibrator (either 3C286 or 3C48)
observed before or after the run. Visibilities taken when the
telescope was driving between source and calibrator are flagged out,
as are those in which one antenna is shadowed by another, or in which
the real or imaginary part is greater than 3.5 times the rms for that
run. (Only a very small fraction of the data is lost by this
amplitude cut, which is done to remove rare interference spikes.) A
map is made of each observation as a check for problems, and the files
are then concatenated. The S-Z signal from a cluster falls very
rapidly with increasing baseline, whereas most radiosources are
unresolved on all our baselines. We therefore use the
1.25--5.4~k$\lambda$ data to estimate the source flux densities and
positions and subtract them from the 0.65--1.25 k$\lambda$ data.

This procedure has been used successfully to map the S-Z effect in more than
10 X-ray-selected clusters (e.g. Saunders 1995); 
in each case the S-Z image has been consistent
with the X-ray data. Very long integrations with the RT have shown no evidence
for any artefacts, and we are confident that this observing process is robust.

\subsection{MS00365 and PG0117+213}

MS00365 is an X-ray-selected quasar at $z = 1.25$ (Stocke et
al. 1991)\nocite{smg91}. It was observed with the RT on 14 occasions
in 1995 June and July, using 0007+171 as a phase calibrator. The map
of all the data had a noise level of 70 $\mu$Jy beam$^{-1}$, and no
features brighter than 240 $\mu$Jy beam$^{-1}$ (3.4 $\sigma$). No
sources were subtracted. The map of baselines shorter than 1.25
k$\lambda$ had a noise level of 178 $\mu$Jy beam$^{-1}$, and no
features brighter than 520 $\mu$Jy beam$^{-1}$ (2.9 $\sigma$).

PG0117+213 is a quasar at $z = 1.49$ with several Ly-$\alpha$ 
absorption systems
(Lanzetta et al. 1995). We observed it with the RT on 8 occasions in 1994 April
and May, with 0149+218 as phase calibrator. No significant features were
apparent on either the 1.25--5.4 k$\lambda$ or 0.65-1.25 k$\lambda$ maps, which
had noise levels of 80 and 180 $\mu$Jy beam$^{-1}$ respectively.

\subsection{PC1643+4631}

The quasar pair PC\,1643+4631\,A \& B was discovered on the basis of
strong emission lines in the Palomar Transit Grism Survey (see
Schneider, Schmidt \& Gunn 1994).  Optical spectra, giving redshifts
of $z=3.79$ and $z=3.83$ respectively, have been published by
Schneider, Schmidt \& Gunn (1991).  Schneider et al. 1991 give $r_{4}$
magnitudes of 20.3, 20.6 for A, B respectively; both are radio quiet.
The quasars lie $198\arcsec$ apart on the sky, corresponding to a
projected distance of 1.3~Mpc (we take $H_{0}=50$
km\,s$^{-1}$Mpc$^{-1}$, $\Omega=1$ and $\Lambda = 0$).  There is a
damped Ly-$\alpha$ absorption system at $z=3.14$ in the spectrum of
quasar~A, making it the subject of many recent observational
programmes (e.g. Hu \& Ridgway 1994).

\begin{figure}[t!]
\plotone{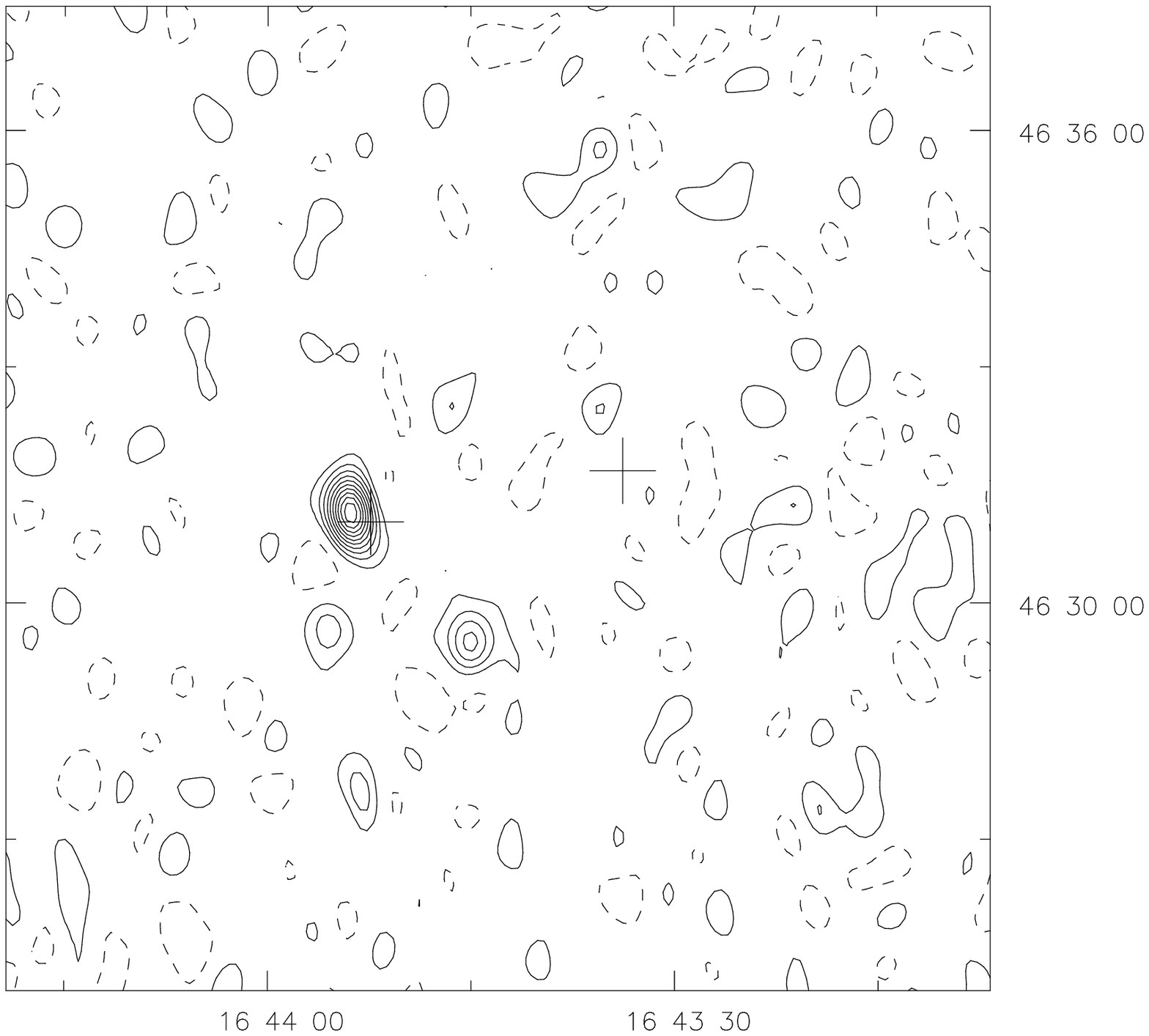}
\caption{PC1643+4631: {\sc clean}ed map of the 1.25--5.4 k$\lambda$ baseline data. The crosses
indicate the positions of quasars A (right) and B (left). Contour levels are
$-50, 50, 100 \dots 500 \mu$Jy, and the beamsize is $30 \arcsec \times 42 \arcsec$. The bright radiosource is {\em not} coincident
with quasar B given the positional errors. Coordinates are B1950.}
\label{1-6cln}
\end{figure}

We observed PC1643+461 on 44 occasions between 1994 March and 1995 June,
pointing the telescope mid-way between the two quasars (at {$\rm 16^h 43^m
43^s +46^{\circ} 31 \arcmin 20 \arcsec$}; all positions are B1950.0). The
phase calibrator was 1624+416, which has a 15-GHz flux density of 1.1~Jy.
Fig.  \ref{1-6cln} shows the {\sc clean}ed map of the 1.25--5.4 k$\lambda$
data. The noise level is 33~$\mu$Jy beam$^{-1}$. Two point sources are evident:
550~$\mu$Jy at {\rm $16^h 43^m 53^s.9 +46^{\circ} 31\arcmin 10\arcsec$}, and
200~$\mu$Jy at {\rm $16^h 43^m 44^s.9 +46^{\circ} 29\arcmin 32\arcsec$}. These
were subtracted from the visibilities and another 1.25--5.4 k$\lambda$ map
made. A third source was found of 150~$\mu$Jy at {\rm $16^h 43^m 55^s.4
+46^{\circ} 29\arcmin 40\arcsec$}; it also was subtracted.

\begin{figure}[t!]
\plotone{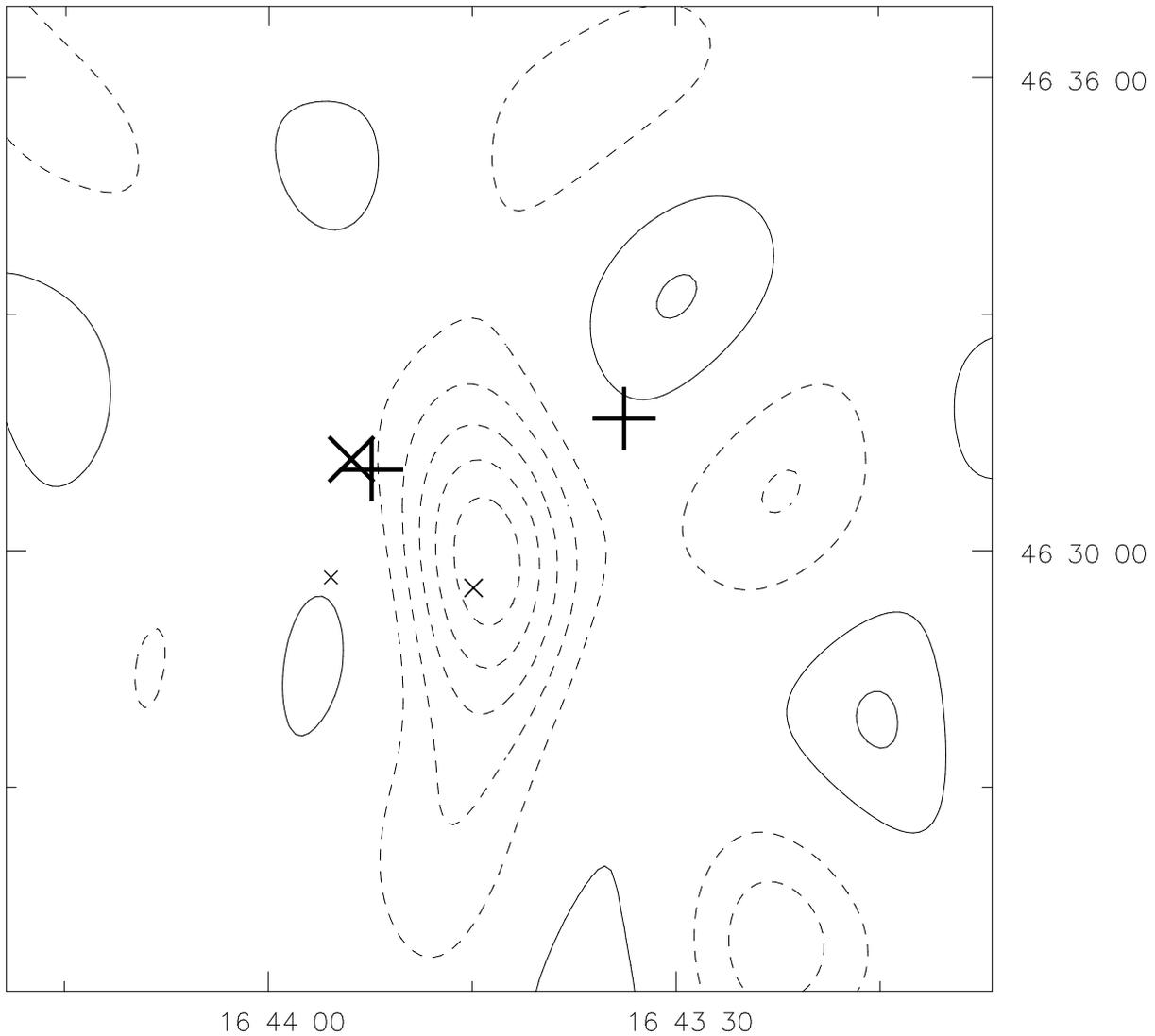}
\caption{PC16433+4631: {\sc clean}ed map of the 0.65--1.25 k$\lambda$
baseline data after source subtraction. The `+' crosses indicate the
positions of quasars A (right) and B (left). The `$\times$' crosses
show the positions of the sources removed; the size of the symbol is
proportional to the removed flux. Contour levels are $-325$ to $+130
\mu$Jy in steps of $65 \; \mu$Jy; dashed contours are negative. Data
from the two different pointings have been added together, weighted
for the envelope beam attenuation. The final map is not corrected for
this attenuation, so the noise level is uniform across the
map. Coordinates are B1950.}
\label{0-1cln}
\end{figure}

We then made a map of the baselines shorter than 1.25~k$\lambda$. This
showed a negative source of $-387 \pm 75 \:\mu$Jy, centred at ${\rm
16^h 43^m 44^s.0 +46^{\circ} 30\arcmin 20\arcsec}$, some $60 \arcsec$
south of the pointing position. Although this is close to the position
of the 200-$\mu$Jy source we subtracted, the magnitude of the negative
source is $> 10$ times the uncertainty in the subtracted flux, and if
the subtracted source were resolved on the short baselines we would
expect to see excess positive, not negative, flux. To check for
possible systematic effects we divided the data into two independent
sets, first by frequency channel and then by time, and re-made the
maps; in each case the divided maps were consistent with the summed
map. We re-observed the field for a further 10 days with the pointing
centre shifted $40\arcsec$ south; after subtracting sources using the
same technique as before, the 0.65--1.25 k$\lambda$ map was consistent
with the previous result, showing a decrement of $-460 \pm 150 \;
\mu$Jy at the same position.  We then made a combined map using the
data from both pointings, taking into account the differing envelope
beam attenuations. This map is shown in Fig. \ref{0-1cln}, which has
been {\sc clean}ed with a restoring beam of $110 \arcsec \times 175
\arcsec$. The peak flux density is $-380 \pm 64 \;\mu$Jy beam$^{-1}$
and the integrated flux is $-410 \; \mu$Jy, i.e. the source is not
significantly extended on this map. The positional accuracy of the
centre of the decrement is roughly the beamsize divided by the
signal-to-noise ratio, i.e. about $20\arcsec \times 30\arcsec$.

\section{X-ray observations of PC1643+4631}

The field of PC1643+4631 was serendipitously observed in a ROSAT PSPC
observation of the cluster A2219. The quasar pair is 52$'$ from the
centre of the PSPC observation, so the effects of vignetting and
degraded point-spread function are severe, but nevertheless this
observation provides a limit to the X-ray flux from the region. In the
total exposure of 11.2 ks, 812 counts were detected within a radius of
$5'$ of the decrement (roughly equal to the size of the local
point-spread function). The backgrounds measured in similar regions of
the detector predict a total of $ 863\pm 30$ counts from this region,
giving a 3-$\sigma$ upper limit of 90 counts above the
background. This corresponds to an unabsorbed flux of $1.7 \times
10^{-16}$ W m$^{-2}$ in the energy range 0.1--2.4~keV
(observed). Assuming a temperature of 5~keV and a redshift of $z=1$,
this corresponds to an X-ray luminosity of $<7\times 10^{37}$ W.

\section{Discussion}

The most conservative assumption is that we have detected an S-Z effect due to
a previously unknown cluster. Primordial CMB anisotropies are expected to have
much smaller amplitudes on arcminute scales.  We have investigated (Lasenby et
al. in preparation) the possibility that the decrement is due to the
Rees-Sciama effect (the change in CMB photon energy on passing through the
gravitational potential of a collapsing object) and it seems very difficult to
produce a significant signal even in extreme cases.

Since the redshift of the cluster is not known and the S-Z decrement is
only detected on a very restricted range of baselines, we have little
information on the angular extent of the decrement. However, the {\em
minimum} magnitude of central decrement is obtained when the cluster
just fills the synthesised beam: if the cluster is smaller than the
beam then there is beam dilution and the observed signal falls; if the
cluster is larger than the beam then the signal is resolved and again
falls. Modelling the cluster as a spherical King model with an angular
dependence of $\Delta T = \Delta T_0 (1 + \theta^2 /
\theta_c^2)^{\frac{1}{2} - \frac{3}{2} \beta}$ with $\beta = 2/3$, the
minimum magnitude of $\Delta T_0$ consistent with our data is $560 \;
\mu$K; the corresponding value of $\theta_{\rm c}$ is about $60
\arcsec$.

We can put a constraint on the minimum redshift of the hot gas from the lack
of observed X-ray emission. In known clusters, an S-Z decrement of this
magnitude would normally be associated with an X-ray luminosity of $\sim
10^{38}$ W. Given the limit in section 4, if this system is similar to those
already known to cause S-Z effects, it must lie at $z \gtrsim 1$. Modelling a
cluster at $z=1$, if we assume $\theta_c = 60''$ and an electron temperature
of $T_e = 5.8 \times 10^7$ K ($\equiv 5$ keV), we get a good fit to the S-Z
data for a central electron density of $3.5 \times 10^3$ m$^{-3}$. Such a
cluster has a gas mass inside a 2-Mpc radius of $3.5 \times 10^{14} \; {\rm
M_{\odot}}$. However, the core radius is 500~kpc, very large by the standards
of nearby clusters. Assuming a core radius of 300~kpc and the same
temperature, we get a best-fit central density of $6.3 \times 10^3$~m$^{-3}$,
and a mass within 2~Mpc of $2.0 \times 10^{14} \; {\rm M_{\odot}}$. In either
case, the total mass (gas plus dark matter) of the cluster is $> 10^{15} {\rm
M_{\odot}}$; this derived mass is also insensitive to the assumed redshift of
the cluster in the interval $4 \gtrsim z \gtrsim 0.4$.  Most ``bottom-up'' models
of structure formation (e.g. cold dark matter, see e.g. Peebles 1993) predict
that large clusters should generally be found at low redshifts, so that
distant clusters massive enough to cause the CMB imprint we observe should be
very rare. It is accordingly now important to search for more clusters via
their CMB imprints. This will provide critical information on the space
densities of massive systems at redshifts in between the two epochs currently
accessible: that at $ z \sim 1000$ obtained from primordial CMB anisotropies
and that close to the present as seen in galaxy surveys.

Further optical and infrared observations and interpretation are presented in
a companion paper (Paper II: Saunders et al. 1996).

\section{Conclusions}

In Ryle-Telescope observations of the field towards the quasar pair
PC1643+4631 A \& B we have obtained a firm detection of a CMB decrement of
$-380 \pm 64~\mu$Jy in a $110 \arcsec \times 175 \arcsec$ beam.  The minimum
magnitude of temperature decrement consistent with these observations is $560
\mu$K, for a spherical King model with $\theta_{\rm c} = 60 \arcsec$. A
serendipitous ROSAT observation places a redshift limit of $z \gtrsim 1$ for a
cluster with an X-ray luminosity of $\sim 7 \times 10^{37}$ W. For a cluster
with a temperature of $\sim 5$ keV, this implies a total mass of $> 10^{15} \;
{\rm M_{\odot}}$, which is only weakly dependant on the assumed temperature
and redshift. These results demonstrate that massive systems too faint to be
detected easily by their emission can readily be seen by their imprints on the
CMB.

\subsection*{Acknowledgments}
We thank the many staff at MRAO who contribute to the running of the Ryle
Telescope, which is supported in part by the PPARC. We thank the referee, Tony
Readhead, for helpful comments. HR ackowledges support from an EU twinning
project and the Netherlands Organization for Scientific Research (NWO).

\end{document}